%Paper: hep-th/9410161
%From: "Connie Jones, University of Rochester" <CONNIE@urhep.pas.rochester.edu>
%Date: Fri, 21 Oct 1994 10:24:48 -0500 (EST)

\magnification=1200
\baselineskip=20pt
\tolerance=100000
\ \
\bigskip
\bigskip

\centerline{Dimensionality as a Perturbation Parameter in the}

\centerline{Generalized Hydrogen Atom}
\vskip 1.5 cm

\centerline{by}
\centerline{C. R. Hagen}
\centerline{Department of Physics and Astronomy}
\centerline{University of Rochester}
\centerline{Rochester, NY  14627}
\vskip 1 cm

\noindent{Abstract}

A recent suggestion has been made that the hydrogen bound state
spectrum should not depend on the number of spatial dimensions.
It is pointed out here that the uncertainty principle implies that
such differences must exist and that a perturbation expansion in
the dimensionality parameter yields a precise quantitative
confirmation of the effect.


\vfil\eject

The solution of the Schrodinger equation for the hydrogen atom
in other than three dimensions has been the object of much study in
recent decades.  In particular the bound state spectrum has been found
[1] in the two dimensional case to be of the form
$$E_n = - {Me^4\over 2\hbar^2(n-1/2)^2} \qquad n = 1,2,...\eqno(1)$$
where $M$ and $e$ are respectively the mass and charge of the electron
and $n$ is the principal quantum number.  Subsequently the result (1)
has been rederived [2-8] in a number of different ways.  The most
striking feature of (1) is perhaps the fact that because of the
occurrence of the $n - 1/2$ factor it predicts a two-dimensional ground
state energy which is four times that of the corresponding
three-dimensional one.

Despite the considerable familiarity achieved by the result (1), it is
far from universally known. Thus, for example, a recent work
 [9] in which (1)
has been derived yet again proposes to \underbar{require} that it reproduce the
three-dimensional result.  Since $n$ is linear in the angular momentum
quantum number $m$, this can be achieved if $m$ (and thus $n$) is
required to be half-integral.  On the other hand this implies the
existence of wave functions which are not single-valued when $m$ assumes
such noninteger values.

Since the literature on this problem is extensive, there is no need here
for one more derivation of the result (1).  Suffice it to say that there
is no real motivation for the half integral angular momentum hypothesis
to bring the two and three dimensional results into agreement.  On the
other hand ref. 9 does pose a somewhat interesting issue in suggesting
that the presence of a third dimension on physical grounds should not
affect the solution.  Such a claim is certainly correct at the classical
level where the three dimensional central field problem can always be
confined to consideration of motion in a plane.

It is clear, however, that quantum mechanical considerations do not allow
this simple picture, which would only be valid if one could
simultaneously require both the momentum and coordinate associated with
the third dimension to vanish.  This, of course, suggests that the
uncertainty principle might profitably be used to clarify the physical
picture.

A highly simplified invocation of the latter could proceed in the
following way.  One can imagine that the third coordinate enters the
problem in a fairly trivial manner by including only its contribution
$p_z^2/2M$ to the Hamiltonian.  This could be a realistic picture for
orbits which are essentially circular and correspond to large radii.  In
this idealization the energy levels can be expected to be raised in the
three dimensional case relative to the two dimensional one by an amount
of the order of $p_z^2/2M$ where the magnitude of $p_z$ is estimated by
the uncertainty principle to be
$$\vert p_z\vert \sim \hbar/a_0$$
where $a_0 = \hbar^2/Me^2$ is the Bohr radius.  This is readily seen to
raise the original two dimensional value by an amount of the order of
$n^{-2}$ Rydberg units, which actually provides a very reasonable
approximation to the energy shift associated with the third dimension.

It is actually possible to improve considerably this crude calculation,
and in fact that is the principal aim of this paper.  Also it is of
interest to note that since Nieto [5] has given the bound state spectrum
of the $N$ dimensional hydrogen atom
$$E_n = - {Me^4\over 2\hbar^2} {1\over [n + 1/2 (N-3)]^2}\>,\eqno(2)$$
it is possible to establish an even more general result.

One begins with the Hamiltonian
$$H = {1\over 2M} p^2 - {e^2\over r}\eqno(3)$$
where
$$\eqalign{p^2 &= \sum^N_{i=1} p_i^2\cr
\noalign{\vskip3pt}
r^2 &= \sum^N_{i=1} x_i^2\cr}$$
and it is assumed that  $N \geq 2$.
The concern here is principally with the additional dimensions
$N > 2$.  Thus one writes
$${1\over r} = (x^2_1 + x^2_2)^{- 1/2} - {1\over 2}(x_1^2 + x_2^2)^{-3/2}
\sum^N_{i=3} x^2_i + \ldots$$
This can be expected to provide a valid basis for an expansion of the $N$
dimensional hydrogen energy levels in terms of the $N=2$ spectrum
provided that one restricts consideration to orbits for which $x_1^2 +
x_2^2$ is large and essentially constant (i.e., circular orbits with $n
\to \infty$).

Thus the Hamiltonian (3) is approximated by
$$H = H_0 + H_1$$
where $H_0$ has the $N=2$ hydrogen atom form and
$$H_1 = \sum^N_{i=3} \left({1\over 2M}p_i^2 + {1\over 2} e^2\langle (x_1^2
+ x_2^2)^{-3/2}\rangle x_i^2\right)\>.$$
Evidently the correction to the ground state $N=2$ energy is simply the
energy associated with $N-2$ harmonic oscillators.  Thus the minimum
correction to the $N=2$ spectrum is given by
$$E_n = - {Me^4\over 2\hbar^2(n- 1/2)^2} + {1\over 2}
(N-2)\hbar\omega\eqno(4)$$
where
$$\omega^2 = {e^2\over M} \langle (x_1^2 + x_2^2)^{-3/2}\rangle\>.$$ From
 the corresponding three dimensional result one finds readily that
for $N=2$ in a state of orbital angular momentum $m$ and principal
quantum number $n$
$$\langle (x_1^2 + x_2^2)^{- 3/2}\rangle_{n,m} = a_0^{-3}[(n-1/2)^3\vert
m \vert (m^2 - 1/4)]^{-1}$$
so that in the most nearly circular orbit states (i.e., $n=\vert m\vert +1$)
$$\omega = {Me^4\over \hbar^3} {1\over (n-1/2)^2} {1\over
[(n-1)(n-3/2)]^{1/2}}\>.\eqno(5)$$

Upon expanding the exact result (2) to first order in the parameter $N-2$
one obtains for the energy shift associated with the $N-2$ additional
dimensions
$$\Delta E_n = {Me^4\over 2\hbar^2} {N-2\over (n-1/2)^3}\>.\eqno(6)$$
On the other hand (4) and (5) yield for this quantity
$$\Delta E_n = {Me^4\over 2\hbar^2} {N-2\over (n-1/2)^2}
{1\over [(n-1)(n-3/2)]^{1/2}}$$
which clearly reproduces (6) in the large $n$ limit.

The calculation presented here has succeeded in reproducing exactly the
minimum (positive) energy contribution associated with each additional
spatial dimension of the generalized hydrogen atom.  This has required
(as expected) that the orbits be large and essentially circular so that
the coefficient of $x_i^2$ in the expansion of the $1/r$ potential can be
made as small and as nearly constant as possible.  As a result of this
fairly elementary exercise one can also better appreciate what might
otherwise be considered a fairly odd mathematical fact -- namely, that the
ground state of hydrogen for $N=2$ is four times that of the $N=3$ case.
Specifically, in the latter case a certain positive amount of energy must
be added (i.e., the energy must be raised) merely to accomplish even a
minimal localization in the third coordinate.

\noindent Acknowledgement

This work was supported in part by U.S. Dept. of Energy grant
DE-FG02-91ER40685.

\noindent References

\item{1.} T. Shibuya and C. E. Wulfram 1965 {\it Amer. J. Phys.} {\bf 33} 570
\item{2.} B. Zaslow and M. E. Zendler 1967 {\it Amer. J. Phys.} {\bf 35} 1118
\item{3.} A. Cisneros and H. Y. McIntosh 1969 {\it J. Math. Phys.} {\bf
10} 277
\item{4.} J. W-K Huang and A. Kozycki 1979 {\it Amer. J. Phys.} {\bf 47} 1005
\item{5.} M. M. Nieto 1979 {\it Amer. J. Phys.} {\bf 47} 1067
\item{6.} G. Q. Hassoun 1981 {\it Amer. J. Phys.} {\bf 49} 143
\item{7.} X. L. Yang, S. H. Guo, F. T.Chan, K. W. Wong, and W. Y. Ching
1991 {\it Phys. Rev. A} {\bf 43} 1186
\item{8.} D. S. Bateman, C. Boyd and B. Dutta-Roy 1992 {\it Amer. J.
Phys.} {\bf 60} 833
\item{9.} V. B. Ho 1994 {\it J. Phys. A: Math. Gen.} {\bf 27} 6237

\vfil\eject
\end